\documentclass{article}
\renewcommand{\theequation}{\arabic{equation}}

\usepackage[dvips]{graphicx}
\usepackage{psfrag}
\usepackage{subfigure}
\begin{document}

\renewcommand{\theequation}{\arabic{equation}}

\begin{center}
{\Large {\bf Spontaneous Symmetry Breakings in }}

{\Large {\bf $Z_2$ Gauge Theories for Doped Quantum Dimer and 
Eight-Vertex Models }}

\vskip 1cm

{\Large  Ikuo Ichinose\footnote{Electric address: ikuo@nitech.ac.jp}
 and Daisuke Yoshioka\footnote{Electric address: daisuke@phys.kyy.nitech.ac.jp}}  

{Department of Applied Physics, Nagoya Institute of Technology,
Nagoya, 466-8555 Japan}

\end{center}
\setcounter{footnote}{0}
\begin{center} 
\begin{bf}
Abstract
\end{bf}
\end{center}
Behavior of doped fermions in $Z_2$ gauge theories for the quantum dimer 
and eight-vertex models is studied.
Fermions carry both charge and spin degrees of freedom. 
In the confinement phase of the $Z_2$ gauge theories, these internal 
symmetries are spontaneously broken and a superconducting or Ne\'el
state appears as the groundstate of the doped fermions.
On the other hand in the 
deconfinement-topologically-ordered state, all symmetries are respected.
>From the view point of the quantum dimer and eight-vertex models, 
this result indicates interplay of the phase structure of the doped
fermions and that of the background dimer
or the eight-vertex model.
At the quantum phase transitions in these systems,
structure of the doped fermion's groundstate and also that of the background
dimer or eight-vertex groundstate both change.
Translational symmetry breaking induces a superconducting or 
antiferromagnetically ordered state of the doped fermions.


\section{Introduction}

Study of the strongly-correlated electron systems is one of the
most interesting problems in the condensed matter physics.
Most of the strongly-correlated electron materials like 
the high-temperature($T$) cuprates, the heavy fermions, etc. 
have various phases like the superconducting state, antiferromagnetic(AF)
Ne\'el state, etc. in very close parameter regions.
This fact indicates that in those materials the charge and spin degrees 
of freedom of electrons correlate and interplay with each other
and as a result the phase transitions occur.
Then it is important to understand quantum phase transitions(QPTs) 
which are difficult to be described by the ordinary Ginzburg-Landau theory 
in terms of a {\em single} order parameter.
Theoretical studies on the QPTs are welcome\cite{sachdev} 
but at present there is 
no unified and transparent understanding of the QPTs which connect 
various phases with {\em different} order parameters.

In this paper we shall investigate the QPTs in which various
order parameters interplay with each other.
In particular we are interested in the quantum dimer and
eight-vertex models with the doped fermions.
Roughly speaking,
the ``background" dimer state can be regarded as f-electrons and 
the doped fermions are electrons in a conduction band from the 
view point of the heavy fermion materials.
As the dimer states can be described by a $Z_2$ gauge theory as it
is well-known\cite{dimer}, the fermion doped system under study is 
described by a 
$Z_2$ Ising gauge theory(IGT) coupled with the fermions.
The doped fermion has {\em both the charge and spin degrees of freedom}.
Then it is interesting to see if there exists an interplay between
background dimer states like valence-bond crystal or liquid and the order
of charge and/or spin degrees of freedom of the doped fermions.

In the elementary particle physics, it is well know that 
confinement of quarks induces spontaneous chiral
symmetry breaking, i.e.,
phase transition of confinement and that of the chiral-symmetry breaking 
occur at the same time.
Then one can expect that similar interplay of different phase 
transitions occurs in the present model.
We shall study this interesting possibility in this paper.

The present paper is organized as follows.
In Sec.2, we shall study the IGT coupled with fermions.
In particular behavior of the
charge and spin degrees of freedom of the fermions is investigated
in strong-coupling regions.
We show that in the even IGT the superconducting state is realized,
whereas in the odd IGT the Ne\'el state appears as the groundstate.
On the other hand in the weak-coupling region, any symmetry breaking 
does not occur.
In Sec.3, we apply the above result to the quantum dimer model on the
square lattice. 
In the valence-bond crystal phase, a dimer or the Ne\'el like state
of the fermion's spin degrees of freedom appears.
In Sec.4, we consider the doped quantum eight-vertex(q8v) model.
In the ordered state of the q8v model, the doped fermions are in the
superconducting phase, whereas in the disordered phase
low-energy excitations are weakly interacting fermions.
Section 5 is devoted for conclusion.


\section{Symmetry breaking in the IGT with fermions}

In this section, we shall first review the IGT and study 
behavior of fermions coupled with IG variables\cite{kogut}.
We use the Hamiltonian formalism on $d$-dimensional hypercubic spatial
lattice.

The IG variables are defined on links of the lattice $(x,\ell)$
where $x$ denotes site and the direction index $\ell=1,\cdots,d$.
Obviously $(x,\ell)=(x+\hat{\ell},-\ell)$ where $\hat{\ell}$ is the
unit vector of the $\ell$-th direction.
Quantum state of the IGT is specified by the
quantum state at each link
which is a linear combination of ``up" and ``down" states;
$c_1 |\uparrow\rangle +c_2 |\downarrow\rangle
=(c_1,c_2)$.
Then the IG operator is given by $\sigma^3_\ell(x)$,
where $\sigma^3_\ell(x)$ is the Pauli spin matrix acting on the
gauge-field configuration $(c_1,c_2)$ at $(x,\ell)$.
Similarly conjugate variable, i.e., the electric field operator, is
given by $\sigma^1_\ell(x)$.
Hamiltonian of the IGT is composed of the electric and magnetic terms,
\begin{equation}
\hat{\cal{H}}_\sigma = \Gamma \sum_{link}{\sigma^1}-\kappa\sum_{pl}
{\sigma^3}{\sigma^3}{\sigma^3}{\sigma^3},
\label{Hsigma}
\end{equation}
where $(link)$ and $(pl)$ denote links and plaquettes,
respectively.
$\Gamma$ and $\kappa$ in $\hat{\cal{H}}_\sigma$ are parameters 
and they are related with the
gauge coupling $g$ as $\Gamma\propto g^2$ and $\kappa \propto 1/g^2$.
The above Hamiltonian is supplemented by the physical-state condition.
In the {\em even}(E) IGT, 
\begin{equation}
\hat{G}_x = \prod_{link \in x}{\sigma^1}; \;\; \;
\hat{G}_x|phys\rangle=|phys\rangle,
\label{phys}
\end{equation}
where ($link \in x$) denotes the links emanating from the site $x$.
The phase structure of the EIGT is well known.
For $d=1$, there is no phase transition and only confinement phase
exists.
On the other hand for $d=2,3$, there is a phase transition from 
confinement to Higgs phase as $\Gamma/\kappa$ decreases.

The odd(O) IGT is defined by the same Hamiltonian (\ref{Hsigma}),
whereas the local constraint is changed to,
\begin{equation}
\hat{G}_x|phys\rangle=-|phys\rangle.
\label{phys2}
\end{equation}
It is easily verified that the above constraints (\ref{phys})and (\ref{phys2})
commute with $\hat{H}_\sigma$.
The OIGT is closely related with the dimer model as we explain
later on\cite{dimer}.
Detailed phase structure of the OIGT is not known yet.

Let us introduce fermions $\psi^\alpha(x)$ $(\alpha=1,2, \cdots, S)$ 
and couple them to the IG field $\sigma^3_\ell(x)$.
Fermion part of the Hamiltonian is given as
\begin{equation}
\hat{\cal H}_\psi=-t \sum_{link,\alpha}
\psi^{\dagger}_\alpha(x)\sigma^3_\ell(x)
\psi^\alpha (x+\hat{\ell})+\mbox{H.c}+M\sum_x 
\psi_{\alpha}^{\dagger}(x)\psi^\alpha(x),
\label{Hpsi}
\end{equation}
where $t$ is the hopping parameter and the ``chemical potential" $M$ controls 
the fermion density $\langle \psi^\dagger \psi \rangle$.
The total Hamiltonian of the system is given by
\begin{equation}
\hat{{\cal H}}_T=\hat{{\cal H}}_\sigma+\hat{{\cal H}}_\psi.
\label{HT}
\end{equation}
By the fermion coupling, the physical-state condition changes to 
\begin{eqnarray}
&& \hat{G}_x\cdot e^{i\pi\sum_\alpha \psi^\dagger(x)\psi(x)}
|phys\rangle=|phys\rangle,
\;\; \mbox{for EIGT} \label{EIGT} \\ 
&& \hat{G}_x\cdot e^{i\pi \sum_\alpha \psi^\dagger(x)\psi(x)}
|phys\rangle=-|phys\rangle,
\;\; \mbox{for OIGT}.
\label{phys3}
\end{eqnarray}

Let us consider the EIGT first.
Effect of the coupled fermions to the phase structure will be discussed in the
following section.
It is shown there that the coupling of the fermions enhances the
deconfinement phase but there still exists the confinement-deconfinement
phase transition.
In the deconfinement phase $\kappa/\Gamma \gg 1$,
fluctuation of the gauge field $\sigma^3_\ell(x)$ is small and we can simply
put $\sigma^3_\ell(x) \sim 1$ (up to irrelevant gauge transformation).
The fermions move almost freely and no spontaneous symmetry breaking(SSB) 
of the internal symmetries occurs.
On the other hand in the confinement phase $\Gamma/\kappa \gg 1$,
one can expect that a SSB occurs as in the most of the gauge theories.
A good example is the SSB of chiral symmetry in the quantum chromodynamics.

In order to investigate the SSB in the present gauge model, we shall
derive a low-energy effective model of fermions in the confinement limit
$\Gamma/\kappa \gg 1$\cite{smit}.
In the leading order of $\Gamma$, the lowest-energy states are specified as 
$\sigma^1_\ell(x)=-1$ for all links
and the fermionic sector is highly degenerate.
In order to derive an effective model for the fermionic sector, we introduce
the projection operator ${\cal P}$ to the subspace $|\psi\rangle_{GS}$ 
satisfying
$\sigma^1_\ell(x)|\psi\rangle_{GS}=-|\psi\rangle_{GS}$ for all links,
and we put ${\cal Q}=1-{\cal P}$.
We start with the eigenvalue problem for the original gauge-fermion system, 
\begin{equation}
\hat{\cal H}_T|\psi\rangle=E |\psi\rangle.
\label{HTeigen}
\end{equation}
By using ${\cal P}$ and ${\cal Q}$, we can derive the following equations,
\begin{eqnarray}
&&(\mathcal{P}\hat{\mathcal{H}}_T\mathcal{P}-E\mathcal{P})\left|\psi\right>=
	-\mathcal{P}\hat{\mathcal{H}}_T\mathcal{Q}\left|\psi\right>\nonumber \\
&&(\mathcal{Q}\hat{\mathcal{H}}_T\mathcal{Q}-E\mathcal{Q})\left|\psi\right>=
	-\mathcal{Q}\hat{\mathcal{H}}_T\mathcal{P}\left|\psi\right>\nonumber \\
&& \mathcal{Q}\left|\psi\right>=
 (E-\mathcal{Q}\hat{\mathcal{H}}_T\mathcal{Q})^{-1}\mathcal{Q}
\hat{\mathcal{H}}_T\mathcal{P}\left|\psi\right>
\label{Eigen}
\end{eqnarray}
>From Eqs.(\ref{Eigen}), we have an eigenvalue equation in the subspace
$\mathcal{P}|\psi\rangle=|\psi\rangle_{GS}$,
\begin{equation}
\left[\mathcal{P}\hat{\mathcal{H}}_T\mathcal{P}+\mathcal{P}
\hat{\mathcal{H}}_T\mathcal{Q}\mathcal{Q}
(E-\mathcal{Q}\hat{\mathcal{H}}_T\mathcal{Q})^{-1}\mathcal{Q}
\hat{\mathcal{H}}_T\mathcal{P}\right]
\left|\psi\right>=E\mathcal{P}\left|\psi\right>.
\end{equation}
Then an effective Hamiltonian $H_e(E)$, which acts on the 
subspace $|\psi\rangle_{GS}$, is given by,
\begin{equation}
H_e(E) = \mathcal{P}\hat{\mathcal{H}}_T\mathcal{P}+
\mathcal{P}\hat{\mathcal{H}}_T\mathcal{Q}\mathcal{Q}
(E-\mathcal{Q}\hat{\mathcal{H}}_T\mathcal{Q})^{-1}\mathcal{Q}
\hat{\mathcal{H}}_T\mathcal{P}.
\label{Heff}
\end{equation}
Explicit form of $H_e(E)$ is obtained in the leading order of 
$1/\Gamma$ as follows,
\begin{eqnarray}
H_e(E) &\sim& -\frac{t^2}{2\Gamma}\sum_{x,\ell}\left[
\psi^{\dagger}_{\alpha}(x)\psi^{\alpha}(x+\hat{\ell})
\psi^{\dagger}_{\beta}(x)\psi^{\beta}(x+\hat{\ell})
\right]\nonumber \\
&& -\frac{t^2}{2\Gamma}\sum_{x,\ell}\left[
\psi^{\dagger}_{\alpha}(x+\hat{\ell})\psi^{\alpha}(x)
\psi^{\dagger}_{\beta}(x+\hat{\ell})\psi^{\beta}(x)
\right]\nonumber \\
&&-\frac{t^2}{2\Gamma}\sum_{x,\ell}\left[
\psi^{\dagger}_{\alpha}(x)\psi^{\alpha}(x+\hat{\ell})
\psi^{\dagger}_{\beta}(x+\hat{\ell})\psi^{\beta}(x)
\right]\nonumber \\
&&-\frac{t^2}{2\Gamma}\sum_{x,\ell}\left[
\psi^{\dagger}_{\alpha}(x+\hat{\ell})\psi^{\alpha}(x)\psi^{\dagger}_{\beta}(x)
\psi^{\beta}(x+\hat{\ell})
\right]
\label{simefchi}
\end{eqnarray}
where we have used 
\begin{equation}
\mathcal{P}
\{{\sigma}^3_{\ell}(x)\}{\cal Q}\{{\sigma}^3_{\ell^{\prime}}(x')\}
\mathcal{P}
=\delta_{\ell,\ell^{\prime}}\delta_{x,x'}\mathcal{P}.
\end{equation}
$H_e(E)$ acts on the fermion sector of $|\psi\rangle_{GS}$
and resolves its degeneracy.

Let us comment on the above low-energy effective model
for the strongly coupled gauge theory.
>From the Hubbard model at the half filling, an antiferromagnetic(AF) Heisenberg
model is derived as a result of the strong {\em on-site repulsion}.
The above effective Hamiltonian $H_e(E)$ (\ref{simefchi}) also contains 
the AF magnetic interactions as we see shortly.
Origin of the AF magnetic interactions is {\em not} the on-site repulsion but 
the strong {\em attractive gauge force}.
One may expect that 
this attractive force may induce other SSBs besides the long-range AF order,
like the superconductivity.
This is the case as we see in later discussion.

The fermion fields satisfy the following canonical anticommutation relations
\begin{eqnarray}
&&\displaystyle{\{\psi_{\alpha}^{\dagger}(x),\psi^{\beta}(y)\}
=\delta^{\alpha}_{\beta} }\delta_{xy} \nonumber \\
&&\displaystyle{\{\psi^{\alpha}(x),\psi^{\beta}(y)\}=0 }.
\label{CCR}
\end{eqnarray}
Then it is useful to define the ``spin" and ``charge operators", 
$Q(x)$'s and $P(x)$'s,
\begin{eqnarray}
Q(x) &=& 
\frac{1}{2}\sum_\alpha 
\left[\psi^{\dagger}_\alpha (x),\psi^\alpha (x)\right]=\sum_\alpha
\left(\psi^{\dagger}_{\alpha}(x)\psi^{\alpha}(x)-{1\over 2}\right) \label{q}\\
Q^{\alpha}_{\beta}(x) &=&
 \frac{1}{2}\left[\psi^{\dagger}_{\beta}(x),\psi^{\alpha}(x)\right]
	-\frac{1}{2}\delta^{\alpha}_{\beta}Q(x) = 
\psi^{\dagger}_{\beta}(x)\psi^{\alpha}(x)
-\frac{1}{2}\delta^{\alpha}_{\beta}-
\frac{1}{2}\delta^{\alpha}_{\beta}Q(x) \label{qab}\\
P_{\alpha\beta}(x) &=&
 \frac{1}{2}\left[\psi^{\dagger}_{\alpha}(x),\psi^{\dagger}_{\beta}(x)\right]
	=\psi^{\dagger}_{\alpha}(x)\psi^{\dagger}_{\beta}(x)\label{pab}\\
\overline{P}_{\alpha\beta}(x)&=&
\frac{1}{2}\left[\psi^{\alpha}(x),\psi^{\beta}(x)\right]=
	\psi^{\alpha}(x)\psi^{\beta}(x). 
\label{pbab}
\end{eqnarray}
Commutation relations of $Q(x)$'s and $P(x)$'s are readily obtained
from Eqs.(\ref{CCR}),
\begin{equation}
 [Q(x),Q(x)]=[P_{\alpha\beta}(x),P_{\alpha\beta}(x)] =
[\overline{P}_{\alpha\beta}(x),\overline{P}_{\alpha\beta}(x)]=0  
\end{equation}
\begin{equation}
[Q^{\alpha}_{\beta}(x),Q^{\gamma}_{\delta}(x)]
=\delta^{\alpha}_{\delta}Q^{\gamma}_{\beta}(x)
	-\delta^{\gamma}_{\beta}Q^{\alpha}_{\delta}(x)
\end{equation}
\begin{equation}
[Q(x),P_{\beta\gamma}(x)]=2P_{\beta\gamma}=-2P_{\gamma\beta}\label{qpcom}
\end{equation}
\begin{equation}
[Q(x),\overline{P}_{\beta\gamma}(x)]=
-2\overline{P}_{\beta\gamma}=2\overline{P}_{\gamma\beta}\label{qpbcom}
\end{equation}
\begin{equation}
[P_{\alpha\beta}(x),\overline{P}_{\gamma\delta}(x)]=
\delta^{\gamma}_{\beta}Q^{\delta}_{\alpha}(x)
		-\delta^{\gamma}_{\alpha}Q^{\delta}_{\beta}(x)
		-\delta^{\delta}_{\alpha}Q^{\gamma}_{\beta}(x)
		-\delta^{\delta}_{\beta}Q^{\gamma}_{\alpha}(x).
\label{ppabcom}
\end{equation}
In terms of $Q(x)$'s and $P(x)$'s, $H_e(E)$ can be written as follows,
\begin{eqnarray}
H_e(E) &=& -\frac{3t^2}{2\Gamma}V
+\frac{t^2}{2\Gamma}\sum_{x,\ell}
[P_{\alpha\beta}(x)\overline{P}_{\alpha\beta}(x+\hat{\ell})
+\overline{P}_{\alpha\beta}(x)P_{\alpha\beta}(x+\hat{\ell})] \nonumber\\
&& +\frac{t^2}{2\Gamma}\sum_{x,\ell}[2Q^{\alpha}_{\beta}(x)
Q^{\beta}_{\alpha}(x+\hat{\ell})+Q(x)Q(x+\hat{\ell})],
\label{HQP}
\end{eqnarray}
where $V$ is the total number of sites of the lattice.

Let us consider the simplest case in which the parameter $\alpha$ takes
two values, i.e., $\alpha=\uparrow, \downarrow$.\footnote{Please
do not confuse the fermion index $\alpha=\uparrow, \downarrow$
with that of the $Z_2$ gauge state.}
Then the states at each site are explicitly given by
\begin{equation}
\left|-1\right>,\quad \psi^{\dagger}_{\uparrow}\left|-1\right>,\quad 
\psi^{\dagger}_{\downarrow}\left|-1\right>,\quad 
\psi^{\dagger}_{\uparrow}\psi^{\dagger}_{\downarrow}\left|-1\right>
\end{equation}
where $|-1\rangle $ is the empty state, $\psi^\alpha |-1\rangle=0$ and
$Q|-1\rangle=-|-1\rangle$.

In the EIGT, only two out of four states $|-1\rangle $ and 
$\psi^{\dagger}_{\uparrow}\psi^{\dagger}_{\downarrow}|-1\rangle$
are the physical state for $\sigma^1_\ell(x)=-1$ and by 
Eq.(\ref{EIGT}).
Then in the Hamiltonian $H_e(E)$, the terms $Q^{\alpha}_{\beta}(x)
Q^{\beta}_{\alpha}(x+\hat{\ell})$ in (\ref{HQP}) do not contribute.
The groundstate of $H_e(E)$ can be written as 
\begin{equation}
|S\rangle=\prod_x\Big[u(x)|-1\rangle_x+v(x)
\psi^{\dagger}_{\uparrow}(x)\psi^{\dagger}_{\downarrow}(x)|-1\rangle_x\Big],
\label{Super}
\end{equation} 
where $u(x)$ and $v(x)$ are complex constants which satisfy 
$|u(x)|^2+|v(x)|^2=1$.
Obviously,
the above state $|S\rangle$ is nothing but the superconducting state.
It is straightforward to calculate the expectation value of the energy
for the state (\ref{Super}), and the lowest-energy state is obtained as
$u(x)=u$ and $v(x)=\epsilon(x)v$ where $\epsilon(x)=1(-1)$ for
the even site(the odd site).
Then the energy $E_S$ is obtained as 
\begin{equation}
E_S = -\frac{3t^2}{2\Gamma}V +
\frac{t^2}{2\Gamma}\sum_{x,\ell}[-4u^2v^2+(-u^2+v^2)^2].
\label{ES}
\end{equation}
On the other hand the density of fermions is
\begin{equation}
\Big\langle \sum_\alpha \psi^\dagger_\alpha \psi^\alpha\Big\rangle=
\langle (Q+1) \rangle=2v^2.
\end{equation}
Therefore at the half filling, $u^2=v^2=1/2$ and $E_S=-{3t^2\over \Gamma}V$.

It is not so difficult to show that the same effective Hamiltonian
(\ref{HQP}) is derived in the OIGT at {\em half-filled} case.
In this case, exactly one fermion resides at each site and the
state $\sigma^1_\ell (x)=-1$ is realized for all links. 
In the single-particle sector of the fermions at each site
$\{\psi^\dagger_\uparrow|-1\rangle,\psi^\dagger_\downarrow|-1\rangle\}$, 
only the last two
terms in $H_e(E)$ (\ref{HQP}) operate and the model reduces to 
the AF Heisenberg model.
At present it is established that the Ne\'el state with the long-range
AF order is the groundstate for $d\ge 2$.
Therefore the groundstate of the doped fermions in the half-filled OIGT
is the AF magnet.

In the following section, we shall study the phase structure of the
doped IGT by the mean-field approximation(MFA) before going into
detailed investigation on the doped dimer and q8v models.

\section{Phase structure of the doped IGT}

\begin{figure}[htbp]
\begin{center}
\includegraphics[scale=1.0]{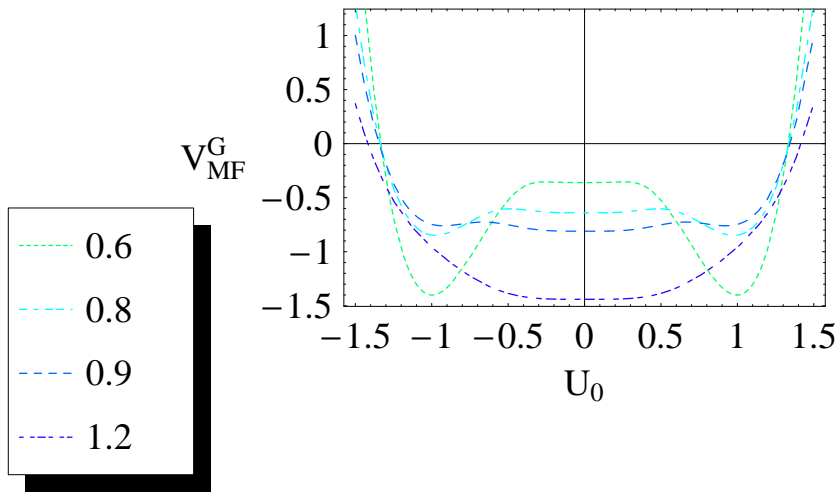}
\end{center}
\caption{Effective potential $V^G_{MF}(U_0)$ of the pure EIGT with the gauge
coupling $g=0.6, \cdots, 1.2$. The result indicates a first-order phase 
transition at $g\simeq 0.9$.}
\label{v_mf_g}
\end{figure}

In this section we shall study the effect of the doped fermions
to the phase structure by the MFA.
We first consider the pure EIGT.
It is not so difficult to derive MF Hamiltonian $\hat{\cal H}_M$ 
from (\ref{Hsigma}).
We formally decompose $\sigma^3_\ell(x)$ into the MF $U_0$ and 
the ``fluctuation" $\delta \sigma^3_\ell(x)$ from it, 
$\sigma^3_\ell(x)=U_0+\delta \sigma^3_\ell(x)$.
The Hamiltonian $\hat{\cal H}_\sigma$ of the IGT is written as 
\begin{eqnarray}
\hat{\cal H}_\sigma &=&\Gamma\sum_{link}
 \sigma^1_\ell(x)+\kappa\Big[-U_0^4N_P-2(d-1)U_0^3\sum_{link}
\delta \sigma^3_\ell(x)
\Big]+O\Big((\delta \sigma^3_\ell (x))^2\Big) \nonumber  \\
&=&\Gamma\sum_{link}\sigma^1_\ell(x)+\kappa\Big[
-U_0^4N_P-2(d-1)U_0^3\sum_{link} 
\sigma^3_\ell(x)+2(d-1)U_0^4N_L\Big] \nonumber \\
&&+O\Big((\delta \sigma^3_\ell (x))^2\Big),
\label{Ssigma}
\end{eqnarray}
where $N_P$ and $N_L$ are the number of plaquettes and links of the 
lattice and for the $d$-dimensional hypercubic lattice $N_L={2\over (d-1)}N_P$.
>From Eq.(\ref{Ssigma}) and by neglecting the terms 
$O\Big((\delta \sigma^3_\ell (x))^2\Big)$, we obtain the MF Hamiltonian
\begin{equation}
\hat{\cal H}^G_M=\Gamma\sum_{link}\sigma^1_\ell(x)+
\kappa\Big[3U_0^4N_P-2(d-1)U_0^3\sum_{link}\sigma^3_\ell(x)\Big].
\label{SM}
\end{equation}
>From Eq.(\ref{SM}), it is straightforward to calculate the
energy of the groundstate as a function of the MF $U_0$.
Here we introduce the gauge coupling constant $g$.
As we stated before $\Gamma=g^2$ and $\kappa=1/g^2$.
The effective potential, i.e, {\em the energy of the groundstate per link},
$V^G_{MF}(U_0)$ is obtained as 
\begin{equation}
 V^G_{MF}(U_0)= {3(d-1) \over 2g^2}U^4_0-\Bigg[\Big({2(d-1)\over g^2}
U^3_0\Big)^2+g^4\Bigg]^{1 \over 2}.
\label{VG}
\end{equation}
In Fig.\ref{v_mf_g} , we show $V^G_{MF}(U_0)$ for various values of 
the gauge coupling $g^2$.
Result obviously shows that there is a first-order phase transition
and at weak coupling $\langle \sigma^3_\ell (x)\rangle \neq 0$, i.e.,
the phase corresponding to the Higgs phase appears\cite{MFA}.
At present for the $3D$ ($d=2$) IGT, it is known that there 
exists a second-order confinement-Higgs phase
transition whereas in the $4D$ ($d=3$) IGT the transition is of 
first-order\cite{PT}.

Effect of the doped fermions is investigated easily by the MFA.
We simply put $\sigma^3_\ell (x) \rightarrow U_0$ in the Hamiltonian 
$\hat{\cal H}_\psi$ (\ref{Hpsi}) to obtain a MF Hamiltonian 
$\hat{\cal H}^\psi_M$,
\begin{equation}
\hat{\cal H}^\psi_M=-tU_0 \sum_{link}\psi^\dagger(x)\psi(x+\hat{\ell})
+\mbox{H.c.}+M\sum_x\psi^\dagger(x)\psi(x).
\label{HMpsi}
\end{equation}
The MF Hamiltonian can be easily diagonalized by the Fourier transformation
of the field operators, 
\begin{equation}
 \hat{\cal H}^\psi_M=\int (d\vec{p})\tilde{\psi}^\dagger(\vec{p})
\Big[-2tU_0(\sum_\ell \cos p_\ell) +M\Big]\tilde{\psi}(\vec{p}).
\label{HP}
\end{equation}
The energy of the groundstate per site is given by
\begin{equation}
V^\psi_{MF}(U_0)=-2t |U_0| \int^{p_F}_0 (d\vec{p})
\Big(\sum_\ell \cos p_\ell\Big),
\label{Vpsi}
\end{equation}
where the Fermi momentum $p_F$ is determined by the density of the 
fermions $\rho_F$, e.g., $p_F=(3\pi^2\rho_F)^{1/3}$ for $d=3$
(and positive $U_0$).
>From (\ref{Vpsi}), it is obvious that $V^\psi_{MF}(U_0)\propto -t |U_0|$
and therefore the fermion doping enhances the condensation of 
$\sigma^3_\ell$, i.e., the deconfinement-Higgs phase.
This result by the MFA, i.e., the linear-$U_0$ dependence of 
$V^\psi_{MF}(U_0)$, probably over estimates the fermion effect.
But we expect that the above result of the {\em enhancement of the 
deconfinement phase} is qualitatively correct as it is observed in 
the various gauge systems.\footnote{Improvement of the simple MFA
was discussed in various places. See for example, Ref.\cite{IM}.}


\section{Doped quantum dimer model}

\begin{figure}[htbp]
\begin{center}
\includegraphics[scale=1.2]{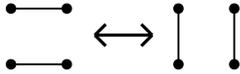}
\end{center}
\caption{Flip of a pair of dimers in a plaquette. Solid lines denote dimer 
and dots are sites of the spatial lattice. }
\label{flipdimer}
\end{figure}

In this section, we shall consider quantum dimer model(QDM)\cite{RK} 
which was introduced as
a model of resonant-valence-bond (RVB) state\cite{RVB}.
We consider the two-dimensional square lattice for
simplicity in the rest of this section.
Quantum states of the model consist of closed-packed and hard-core
dimers.
Hamiltonian of the quantum dimer model is given as
\begin{equation}
H_{dimer}=\sum_{plaquette}\Big[ -q{\cal F}_P+p{\cal V}_P\Big],
\label{Hdimer}
\end{equation}
where ${\cal F}_P$ is a flip term defined on each plaquette 
(see Fig \ref{flipdimer}.)
and ${\cal V}_P$ is the potential term which takes unity for
flipable dimer state in the plaquette and otherwise vanishing.
$q$ and $p$ are parameters of the model.
At the Rokhsar and Kivelson(RK) point $q=p$, the groundstate is given 
by the summation of all
dimer configurations with equal weight; the valence-bond liquid state.

In the previous section we introduced the OIGT.
The OIGT can be regarded as a kind of the QDM.
In the OIGT description of the QDM, the state $\sigma^1_\ell(x)=-1$ on the link
$(x,\ell)$ corresponds to the empty state, whereas 
that of $\sigma^1_\ell(x)=1$ is regarded as the state occupied by a
dimer.
Then the physical-state condition of the {\em undoped} OIGT
with the local constraint
(\ref{phys2}) requires that one or three dimers emanate from
each site.
For large positive $\Gamma$ in the Hamiltonian (\ref{Hsigma}), 
the space of states consists of closed-packed and hard-core
dimers on the square lattice.
On the other hand, the second term of Eq.(\ref{Hsigma}) works as the
flip term of two dimers that reside on a single plaquette
(see Fig.\ref{flipdimer}).

At present it is known that in most of the parameter
region the columnar or staggered ordered state,
which breaks the translational symmetry, is the groundstate of
the undoped QDM $H_{dimer}$ (\ref{Hdimer}) on the square lattice\cite{GSdimer},
whereas at the RK point the translationally symmetric state
is realized\cite{dimer,decRK,MS}.
However by the doping, a liquid state of VB may
appear as the groundstate in a finite parameter region\cite{SV}.

In the previous section we considered the half-filled case of the doped OIGT.
Here we focus on the lightly doped QDM assuming that a valence bond
crystal like the columnar state is still the groundstate.
Let us consider a two-fermion doped QDM first.
A pair of fermions doped in the columnar state reside on the
nearest-neighbor(NN) sites and a dimer on the link connecting
the two sites disappears.
This state is the lowest-energy state.
Similarly when $2n$ fermions are doped into the QDM, there are various
configurations in which $n$ dimers disappear
by the doping. 
Effective Hamiltonian of the doped fermions can be derived in the
framework of the OIGT and the Hamiltonian of the AF Heisenberg model
is obtained (up to an irrelevant constant term),
\begin{equation}
H_e^{dimer}=\frac{t^2}{\Gamma}\sum_{x,\ell}\Big(2Q^{\alpha}_{\beta}(x)
Q^{\beta}_{\alpha}(x+\hat{\ell})+Q(x)Q(x+\hat{\ell})\Big),
\label{HD}
\end{equation}
where the summation is performed over the links $(x,{\ell})$ 
{\em connecting sites of doped fermions}.

\begin{figure}[htbp]
\begin{center}
\subfigure[Columnar]{\includegraphics[scale=1.0]{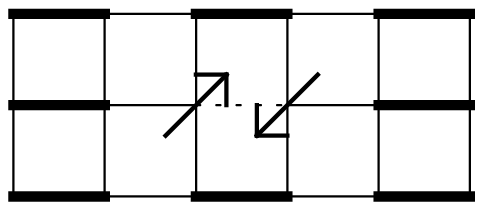}\label{first_doped_col}}
\hspace*{10mm}
\subfigure[Staggered]{\includegraphics[scale=1.0]{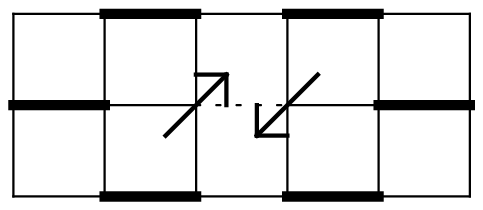}\label{first_doped_sta}}\\
\subfigure[Mixed 1]{\includegraphics[scale=1.0]{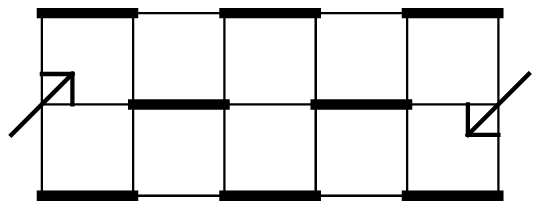}\label{col__sta_mix01}}
\hspace*{10mm}
\subfigure[Mixed 2]{\includegraphics[scale=1.0]{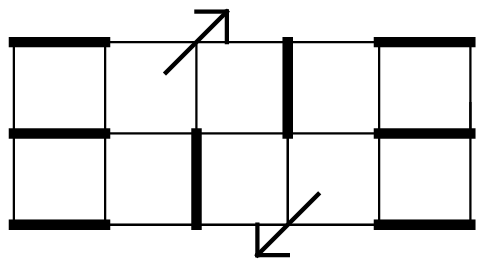}\label{col_sta_mix02}}
\end{center}
\caption{Configurations of fermions in dimer backgrounds: Heavy lines denote 
dimer. Up(down) arrow denotes spin-up(down) state of the doped fermion.}
\label{firstdoped}
\end{figure}

For an isolated pair of fermions, it is easily to obtain the 
wavefunction.
In particular in the case $S=2$, i.e., $\alpha=\uparrow,\downarrow$,
they form a spin-singlet state,
\begin{equation}
\Big(\psi^\dagger_\uparrow (x)\psi^\dagger_\downarrow(x+\hat{\ell})
-\psi^\dagger_\downarrow (x)\psi^\dagger_\uparrow(x+\hat{\ell})
\Big)|-1\rangle_x|-1\rangle_{x+\hat{\ell}},
\label{SSP}
\end{equation}
and the energy of the spin-singlet pair is estimated from (\ref{HD}) as 
$E_{SSP}=-{3t^2\over \Gamma}$.
Two fermions can be put on two sites which separate more than one lattice
spacing if the columnar-staggered-mixed configuration of dimers is
considered (see Fig.\ref{firstdoped}).
However those states have higher energy compared with the spin-singlet pair
state (\ref{SSP}) regardless of their spin configurations as seen 
from $H_e^{dimer}$.
In this way, we can expect that lightly fermion-doped dimer system has
the groundstate with {\em inhomogeneous fermion density}, i.e., high and low
fermion density regions appear.
In the high density region, the Ne\'el-like state for the fermion's internal
degrees of freedom appears or a deconfinement-VB-liquid 
phase is realized.
On the other hand in the low density region, a closed-packed and hard-core
dimer state appears.
Numerical studies are required in order to obtain detailed quantitative
phase diagram of the doped dimer systems.


\section{Doped quantum eight-vertex model as IGT}
\subsection{Eight-vertex model: Short review}
\begin{figure}[htbp]
\begin{center}
\includegraphics[scale=1.0]{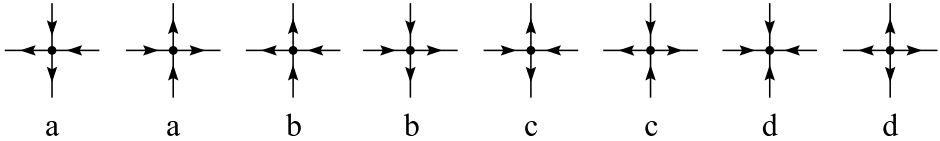}
\end{center}
\caption{Vertex configurations: There are eight vertices in 
eight-vertex model. Their Boltzmann weight is given by $a,\cdots, d$.}
\label{8v_conf}
\end{figure}

In this section we shall investigate the behavior of doped
fermions in the quantum eight-vertex(q8v) model\cite{q8v}.
Classical 8v model was studied by Baxter very intensively\cite{baxter}.
Dynamical variables are ``arrows" sitting on each link of the
square lattice, and then at each site four arrows meet; i.e., vertex.
At each vertex, the numbers of incoming and outgoing arrows are
both restricted to even, and therefore there are eight types of 
vertices, see Fig.\ref{8v_conf}.
For each vertex, the Boltzmann weight $w_i\;(i=1,2,\cdots,8)$ is asigned 
and total Boltzmann weight for a configuration is simply given by the
product of $w_i$ of each vertex composing that configuration.
We choose the weight $w_i$ as follows,
\begin{eqnarray}
&& w_1=w_2=a, \;\; w_3=w_4=b, \nonumber \\
&& w_5=w_6=c, \;\; w_7=w_8=d.
\label{abcd}
\end{eqnarray}
Hereafter, $a,\cdots, d$ also denote type of vertices.
Then the partition function of classical 8v model is given as
\begin{equation}
Z_{c8v}(a,b,c,d)=\sum_{conf's}a^{v_a}b^{v_b}c^{v_c}d^{v_d},
\label{Zc8v}
\end{equation}
where $v_a$ is the number of the $a$-type vertex in the configuration, etc.

Basis of the Hilbert space of the q8v model consists of the 
configuration space of the classical 8v model.
Different configurations of vertices are orthogonal with each other and
the norm of each configuration of vertices is normalized as unity.
Hamiltonian is composed of a flip term and a potential term.
The q8v model can be described by an IGT, which is the subject of the
following subsection.

\subsection{IGT for q8v model}
 
As explained in the previous subsection, state of the q8v model
is specified by assigning arrow for each link of the two-dimensional
square lattice.
We call the states of the up and right arrow spin-up state whereas 
the states of the down and left arrow spin-down state.
In order to construct physical operators, let us introduce the Pauli
spin matrices for each link $(x,\ell)$ $(\ell=1, 2)$.
Let the above up-spin and down-spin states on the link $(x,\ell)$ 
be eigenstates of the operator $\sigma^1_\ell(x)$ with eigenvalue
$\pm 1$, respectively.
>From the restriction on the configurations of the classical 8v model,
the physical state of the q8v model must be satisfied the following 
local constraint,
\begin{equation}
\sigma^1_1(x)\sigma^1_1(x-\hat{1})\sigma^1_2(x)\sigma^1_2(x-\hat{2})
|phys\rangle=|phys\rangle, \;\; \mbox{for all} \;x.
\label{const8v}
\end{equation}
The constraint (\ref{const8v}) is nothing but the gauge-invariant
condition in the IGT.
Flip operation works on arrows on four links forming a plaquette,
and then the flip term of the Hamiltonian is given as 
\begin{equation}
H_{flip}=-\sum_{pl}\sigma^3_1(x)\sigma^3_1(x+\hat{2})
\sigma^3_2(x)\sigma^3_2(x+\hat{1}).
\label{flip8v}
\end{equation}
It is obvious that the above Hamiltonian (\ref{flip8v}) is the magnetic term
of the IGT and commutes with the constraint (\ref{const8v}).

If there are no terms in the Hamiltonian besides the flip term (\ref{flip8v}),
the groundstate is the summation of all the states in the configuration space
with equal weight, $a=b=c=d$.
This state is nothing but Kitaev's state which may play an important role
in quantum tori codes\cite{kitaev}.
In order to give nontrivial weight for states in the q8v model,
we shall introduce a potential term.
To this end we consider the following operators that distinguish
the verices $a,\cdots, d$,
\begin{eqnarray}
&&S_a(x)={1\over 4}\Big(\sigma^1_1(x)+\sigma^1_2(x)+\sigma^1_1(x-\hat{1})
+\sigma^1_2(x-\hat{2})\Big),   \nonumber  \\
&& S_b(x)={1\over 4}\Big(\sigma^1_1(x)-\sigma^1_2(x)+\sigma^1_1(x-\hat{1})
-\sigma^1_2(x-\hat{2})\Big),   \nonumber  \\
&&S_c(x)={1\over 4}\Big(\sigma^1_1(x)-\sigma^1_2(x)-\sigma^1_1(x-\hat{1})
+\sigma^1_2(x-\hat{2})\Big),   \nonumber  \\
&&S_d(x)={1\over 4}\Big(\sigma^1_1(x)+\sigma^1_2(x)-\sigma^1_1(x-\hat{1})
-\sigma^1_2(x-\hat{2})\Big).  
\label{Ss}
\end{eqnarray}
Operation of the above four operators is summarized in Table \ref{table01_8v}.
>From the Table1, it is obvious that all eight vertices are
distinguished by $S_a,\cdots,S_d$. 
Projection operators can be easily obtained,
\begin{equation}
{\cal P}_i=S^2_i, \;\; i=a,b,c,d,
\label{Pj}
\end{equation}
and after some calculation
\begin{equation}
{\cal P}_a+{\cal P}_b+{\cal P}_c+{\cal P}_d={\cal I}.
\end{equation}
\begin{table}[htbp]
\caption{Relation between operator $S_i$ and eight-vertex configurations} 
\label{table01_8v}
\begin{center}
\normalsize
\begin{tabular}{|l|r|r|r|r|r|r|r|r|} \hline
 & \includegraphics[trim=0 8 0 0,scale=0.7]{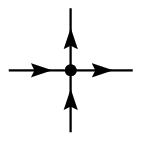} 
 & \includegraphics[trim=0 8 0 0,scale=0.7]{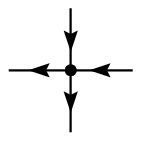}
 & \includegraphics[trim=0 8 0 0,scale=0.7]{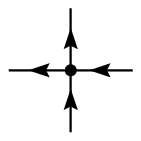} 
 & \includegraphics[trim=0 8 0 0,scale=0.7]{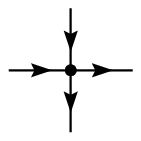} 
 & \includegraphics[trim=0 8 0 0,scale=0.7]{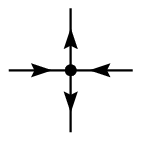} 
 & \includegraphics[trim=0 8 0 0,scale=0.7]{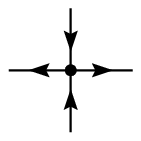}
 & \includegraphics[trim=0 8 0 0,scale=0.7]{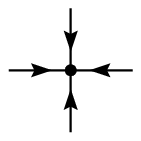}
 & \includegraphics[trim=0 8 0 0,scale=0.7]{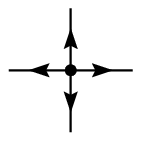}\\ \hline
$S_a$ & 1 & -1 & 0 & 0 & 0 & 0 & 0 & 0 \\ \hline
$S_b$ & 0 & 0 & -1 & 1 & 0 & 0 & 0 & 0 \\ \hline
$S_c$ & 0 & 0 & 0 & 0 &-1& 1 & 0 & 0 \\ \hline
$S_d$ & 0 & 0 & 0 & 0 & 0 & 0 &-1& 1\\ \hline
\end{tabular}
\end{center}
\end{table} %

\begin{figure}[htbp]
 \psfrag{c}[B][B][1.5][0]{$\mathrm{c^2}$}
 \psfrag{d}[B][B][1.5][0]{$\mathrm{d^2}$}
\begin{center}
\includegraphics[scale=0.8]{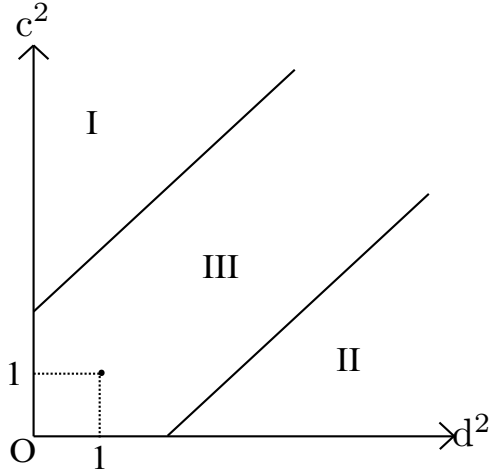}
\end{center}
\caption{Phase Diagram of q8v model. Regions $I$ and $II$ are ordered-solid
phase whereas region $III$ corresponds to disordered-liquid phase.
Doped fermions are in supercondcting phase in region $I$ and $II$.
Region $III$ are normal phase of doped fermions.}
\label{q8v-pd}
\end{figure}

By using the projection operators ${\cal P}_i$ $(i=a,b,c,d)$, 
we can construct a potential term which assigns weight for each pure state 
just like the partition function of the
classical 8v model (\ref{Zc8v}).
To this end, let us see how vertices interchange with each other
under the plaquette flip.
For example the $d$-type vertex on left-down or right-up site(even site) of a 
flip plaquette changes to the $a$-type vertex.
On the other hand, the $d$-type vertex on right-down or left-up site (odd site)
changes to the $b$-type vertex.
These are summarized in Table.\ref{table02_8v}.
We can give the potential term\cite{q8v},
\begin{eqnarray}
H_V&=& \sum_x\Big({d\over a}{\cal P}_a(x)+{c\over b}{\cal P}_b(x)
+{b\over c}{\cal P}_c(x)+{a\over d}{\cal P}_d(x)\Big) \nonumber \\
&&\times \Big({c\over a}{\cal P}_a(x+\hat{1})+
{d\over b}{\cal P}_b(x+\hat{1})
+{a\over c}{\cal P}_c(x+\hat{1})+
{b\over d}{\cal P}_d(x+\hat{1})\Big) \nonumber \\
&&\times \Big({c\over a}{\cal P}_a(x+\hat{2})+
{d\over b}{\cal P}_b(x+\hat{2})
+{a\over c}{\cal P}_c(x+\hat{2})+
{b\over d}{\cal P}_d(x+\hat{2})\Big) \nonumber \\
&& \times\Big({d\over a}{\cal P}_a(x+\hat{1}+\hat{2})+
{c\over b}{\cal P}_b(x+\hat{1}+\hat{2})
+{b\over c}{\cal P}_c(x+\hat{1}+\hat{2})+
{a\over d}{\cal P}_d(x+\hat{1}+\hat{2})\Big).
\label{pot8v}
\end{eqnarray} 
The total Hamiltonian of the q8v model is then given by
\begin{equation}
\hat{\cal H}_{q8v}=H_{flip}+H_V.
\label{HTq8v}
\end{equation}
It is not so difficult to show that weight of the configuration
with $v_a$ $a$-type vertices etc. is given by $a^{v_a}b^{v_b}
c^{v_c}d^{v_d}$ in the groundstate wavefunction $|GS(a,b,c,d)\rangle$.
Then the partition function of the {\em q8v model} at $T=0$ is
related with the {\em c8v model} as follows,
\begin{equation}
Z_{q8v}(a,b,c,d)=Z_{c8v}(a^2,b^2,c^2,d^2).
\label{ZZ}
\end{equation}
%
\begin{table}[htbp]
\caption{Relation between flip operation and eight-vertex configurations:
Cross symbol and circled-cross symbol denote $a$-type and $b$-type vertices,
respectively.
Similarly, dot symbol is $c$-type, and triangle symbol is $d$-type vertices. 
Relation between types of vertices and symbols is summarized in the left table.
The right table shows how vertices interchange with each other 
under plaquette-flip operation.} 
\label{table02_8v}
\begin{center}
\begin{tabular}{|l|l|l|l|}
\hline
a & b & c & d \\
\hline
\includegraphics{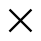} &\includegraphics{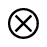}  &
\includegraphics{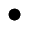}  &\includegraphics{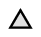}  \\
\hline
\end{tabular}
\hspace{1cm}
\begin{tabular}{|c|c|}
\hline
vertex type & before and after flipping \\
\hline
c & \includegraphics{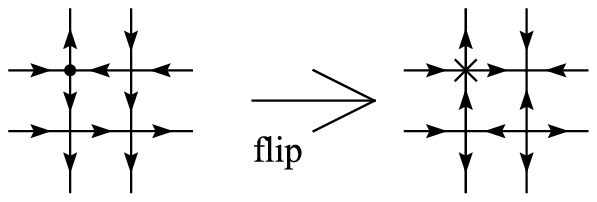} \\
\cline{2-2}
 &  \includegraphics{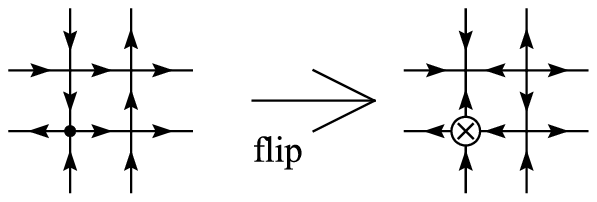}\\
\hline
d & \includegraphics{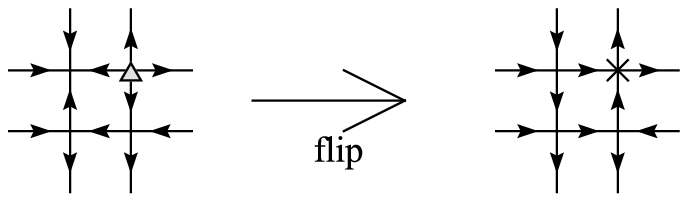}\\
\cline{2-2}
 & \includegraphics{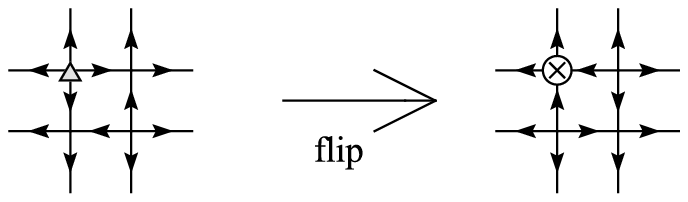} \\
\hline
\end{tabular}
\end{center}
\end{table}
Phase structure of the classical 8v model is well-known, and from 
the relation (\ref{ZZ}) it gives that of the q8v model at $T=0$.
Fig.\ref{q8v-pd} shows the phase diagram of the q8v 
model\cite{baxter,q8v}.
We put $a=b=1$ for simplicity.
The region $I$ ($II$), which is given by $c^2>d^2+2$ ($d^2>c^2+2$), 
is an ordered phase where the $c$-type vertices ($d$-type vertices) dominate.
On the other hand the region $III$, $-2<c^2-d^2<2$ is a disordered phase
and various states including all vertices appear in the wavefunction of 
the groundstate\footnote{In certain limits of the parameter space
like $c^2=d^2\rightarrow \infty$, only $c$ and/or $d$ vertices appear.
See later discussion.}.
In particular at Kitaev's point $a=b=c=d=1$, all configurations appear
with the equal weight.

Let us consider fermion doping in the q8v model.
In the region $III$(the disordered phase), doped fermions move
almost freely and no SSB occurs.
On the other hand in the ordered phases $I$ and $II$, strong interactions
work on doped fermions and we expect that some kind of SSB of the internal
symmetry of fermions takes place there.
This is the subject of the following subsection.


\subsection{The limit $c^2\rightarrow \infty$ with finite $d^2$}

\begin{figure}[htbp]
\begin{center}
\subfigure[$c^2\to\infty$ configuration of q8v model.]
{\includegraphics[scale=1.0]{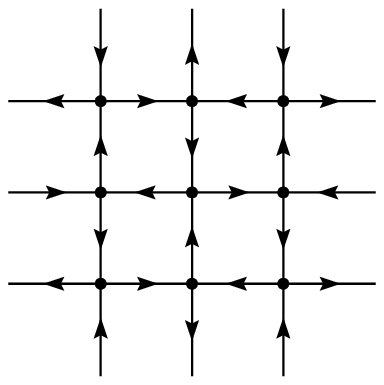}\label{ctoinf}}
\hspace*{1.5cm}
\subfigure[Doped fermions on the same site
in the $c^2\to\infty$ configuration.]
{\includegraphics[scale=1.0]{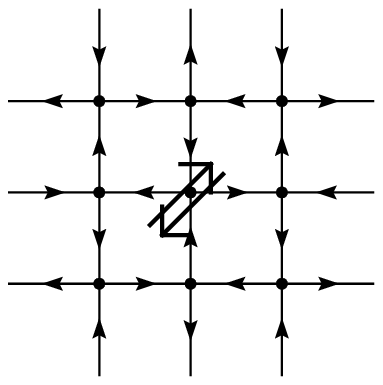}\label{dope_ctoinf1}}
\hspace*{1.5cm}
\subfigure[Doped fermions in the $c^2\to\infty$ configuration.]
{\includegraphics[scale=1.0]{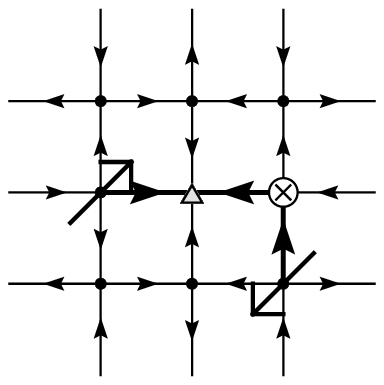}\label{dope_ctoinf}}
\end{center}
\caption{$c^2\to\infty$ configurations.}
\label{ctoinf__}
\end{figure}

We shall consider the specific limit $c^2\rightarrow \infty$ 
for investigating behavior of the doped fermions in the ordered phase.
In this limit, the undoped groundstate 
$|GS;c\rangle=|GS(a,b,c,d)\rangle_{(c^2\rightarrow \infty)}$ satisfies
\begin{equation}
{\cal P}_c(x)|GS;c\rangle=|GS;c\rangle, \;\; \mbox{for all $x$}
\label{GSc}
\end{equation}
and it is shown in Fig \ref{ctoinf}.
Let us dope fermions in the 8v background.
Fermion part of the Hamiltonian is the same with that of the IGT
Eq.(\ref{Hpsi})
and the total Hamiltonian of the doped q8v model is given by
\begin{equation}
\hat{\mathcal{H}}=\hat{\mathcal{H}}_{q8v}+\hat{\mathcal{H}}_\psi.
\end{equation}
The constraint on the state (\ref{const8v}) changes to 
\begin{equation}
\sigma^1_1(x)\sigma^1_1(x-\hat{1})\sigma^1_2(x)\sigma^1_2(x-\hat{2})
\cdot e^{i\pi \sum \psi^\dagger(x) \psi(x)}|phys\rangle=|phys\rangle.
\label{constdoped8v}
\end{equation}

In the ordered phase, the gauge variables $\sigma^1_\ell (x)$'s
take a definite value at each link.
This means that the variables $\sigma^3_\ell (x)$ fluctuate
very strongly, i.e., the {\em ordered phase} of the q8v model is nothing
but the {\em confinement phase} of the IGT.
This fact is pictorially observed for the limit $c^2 \rightarrow \infty$.
In this case, the {\em undoped} system is full of $c$-type vertices 
(see Fig \ref{ctoinf}.).
Then let us dope two fermions into the state $|GS;c\rangle$.
In the lowest-energy state (i.e., the highest-weight state), the two fermions
reside on the same site if their internal quantum number $\alpha$ are
different whereas the fermions of the same quantum number reside 
the nearest-neighbour (NN) site.
In the former case, the gauge-field configuration is the same with 
the undoped state $|GS;c\rangle$.
On the other hand in the latter case, new types of vertex appear at
the sites occupied with the fermions as dictated by the local constraint 
(\ref{constdoped8v}).
Then energy of the resultant gauge configuration is higher than 
that of $|GS;c\rangle$ by an amount $c^4$, which is directly
calculated from $H_V$.
When one of fermions in a pair hops from its original position to
a site with a distance $L$ (on the lattice), $L$ vertices change
from the $c$-type vertices to the other ones because the hopping term
of the fermion is accompanying the gauge operator $\sigma^3_\ell(x)$ 
which flips the arrow on the link $(x,\ell)$.
Then a separated pair of fermions cost energy $\sim Lc^4$.
This means nothing but the confinement of the doped fermions.
See Fig.\ref{ctoinf__}.

Effective Hamiltonian of fermionic degrees of freedom in the 
ordered phase can be derived
as in the previous discussion on the strongly-coupled IGT.
For states with pairs of fermions residing the same sites like 
Eq.(\ref{Super}),
\begin{equation}
H_e^{q8v}(E) \sim \mathcal{P}_c\hat{\mathcal{H}}\mathcal{P}_c+
\mathcal{P}_c\hat{\mathcal{H}}\mathcal{Q}_c\mathcal{Q}_c
(E-\mathcal{Q}_c\hat{\mathcal{H}}\mathcal{Q}_c)^{-1}
\mathcal{Q}_c\hat{\mathcal{H}}\mathcal{P}_c,
\end{equation}
where $(\hat{\mathcal{H}}_{q8v}+\hat{\mathcal{H}}_\psi)
|\psi\rangle=E|\psi\rangle$, ${\cal P}_c=\prod_x{\cal P}_c(x)$
and ${\cal Q}_c=1-{\cal P}_c$.
In the leading order of $c^2$,
\begin{eqnarray}
H_{e}^{q8v}(E) &=& 
-\frac{16}{c^4}\mathcal{P}_c\hat{\mathcal{H}}\mathcal{Q}_c
\hat{\mathcal{H}}_\psi\mathcal{P}_c,  \nonumber \\
&=& \frac{16t^2}{c^4}\sum_{x,\ell}
[P_{\alpha\beta}(x)\overline{P}_{\alpha\beta}(x+\hat{\ell})
+\overline{P}_{\alpha\beta}(x)P_{\alpha\beta}(x+\hat{\ell})].
\label{HEq8v}
\end{eqnarray}
>From (\ref{HEq8v}) and the discussion in Sec.2, it is obvious that 
the {\em superconducting phase} of the fermions is realized in the 
{\em ordered phase} of the doped q8v model.
On the other hand in the disordered phase, no SSBs occur.
We can expect that the phase boundary of the order-disorder phase 
transition of the background q8v system and that of the superconductivity
of the doped fermions coincide, though the region of the deconfinement 
phase (i.e., the disordered phase with the translational symmetry) 
is enlarged by the fermion doping as in the EIGT.
The phenomenon that the confinement induces  SSB of internal symmetry
is generally observed in various gauge theories.

\subsection{The limit $c^2=d^2\rightarrow \infty$}

\begin{figure}[htbp]
\begin{center}
\subfigure[Columnar]{\includegraphics[scale=1.0]{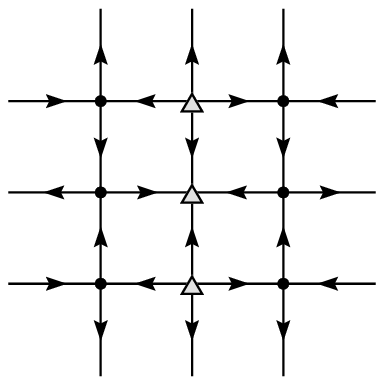}\label{cdtoinf_co}}
\hspace*{10mm}
\subfigure[Staggered]{\includegraphics[scale=1.0]{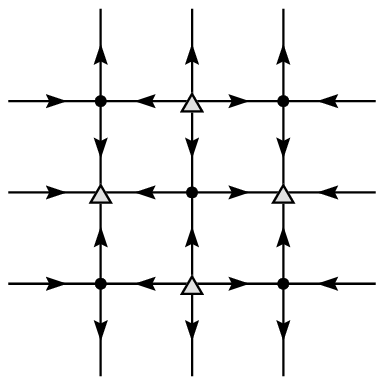}\label{cdtoinf_st}}
\end{center}
\caption{$c^2=d^2 \to \infty$ configurations of q8v model.}
\label{cdtoinf}
\end{figure}

It is interesting and also instructive  to see how the doped
fermions behave in the topologically ordered phase $III$.
To this end we consider the limit $c^2=d^2\rightarrow \infty$ with
keeping $a=b=1$.
We set $c=d=K$ and take the limit $K\rightarrow \infty$.
In this limit, the potential term $H_V$ tends to
\begin{eqnarray}
H_V &=& K^4\sum_{x}
(\mathcal{P}_a(x)+\mathcal{P}_b(x))
(\mathcal{P}_a(x+a_1)+\mathcal{P}_b(x+a_1)) \nonumber \\
&& \times (\mathcal{P}_a(x+a_2)+\mathcal{P}_b(x+a_2))
(\mathcal{P}_a(x+a_1+a_2)+\mathcal{P}_b(x+a_1+a_2)).
\label{HVK}
\end{eqnarray}
>From (\ref{HVK}), the states of the lowest(vanishing) eigenvalue of $H_V$,
$|GS;cd\rangle$, 
are specified by the projection operator $\mathcal{P}_{cd}(x)$,
\begin{equation}
\mathcal{P}_{cd}(x) = \mathcal{P}_c(x) + \mathcal{P}_{d}(x),
\label{Pcd}
\end{equation}
\begin{equation}
\mathcal{P}_{cd}(x)|GS;cd\rangle=|GS;cd\rangle, \;\; \mbox{for all $x$}.
\label{GScd}
\end{equation}
There are various states, besides $|GS;c\rangle$ and $|GS;d\rangle$, 
which satisfy the condition (\ref{GScd}).
As the flip operator $H_{flip}$ in (\ref{flip8v}) is applied on the
$|GS;cd\rangle$, the $a$ and/or $b$ vertices appear.
Therefore in the limit $K \rightarrow \infty$, all various states 
in $|GS;cd\rangle$ are independent and degenerate.
Among them, some symmetric configurations are
depicted in Fig.\ref{cdtoinf_co} and \ref{cdtoinf_st}, i.e., 
the {\em c-d columnar} and {\em c-d staggered} configurations, respectively.

\begin{figure}[htbp]
\begin{center}
\subfigure[Doped fermions in $|GS;c\rangle$]
{\includegraphics[scale=1.0]{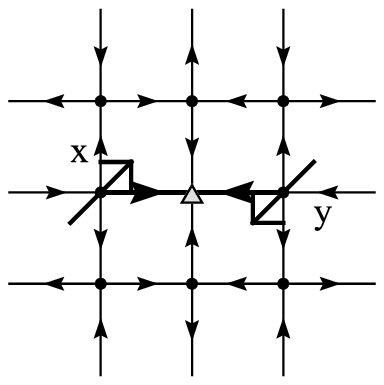}\label{ctoinfc}}
\hspace*{10mm}
\subfigure[Doped fermions in $|GS;d\rangle$]
{\includegraphics[scale=1.0]{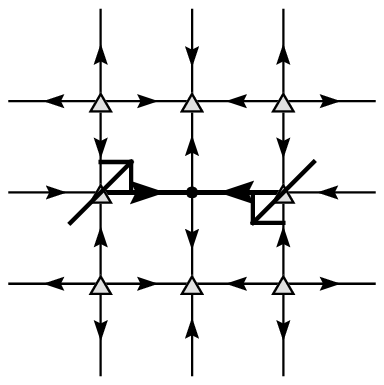}\label{dtoinfc}}\\
\subfigure[Doped fermions in $(c-d)$ columnar state]
{\includegraphics[scale=1.0]{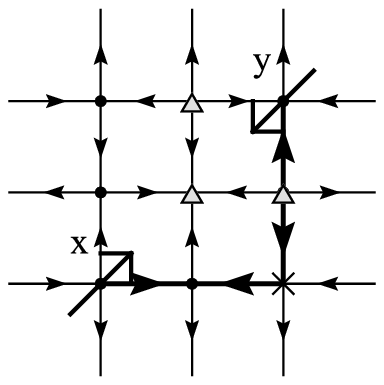}\label{cdtoinfco}}
\hspace*{10mm}
\subfigure[Doped fermions in $(c-d)$ staggered state]
{\includegraphics[scale=1.0]{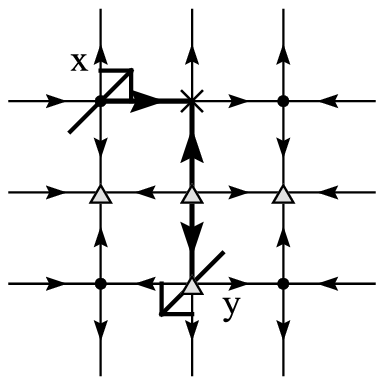}\label{cdtoinfst}}
\end{center}
\caption{Fermion hopping and change of background 8v configurations.
Fermion hopping in one-dimensional straightlines 
changes the $c$-vertices to the $d$-vertices.
In the limit $c^2=d^2\rightarrow \infty$, resultant configurations
of the 8v model have the same energy with the original ones.
Therefore doped fermions move freely in the horizontal and 
vertical lines.
}
\label{4states}
\end{figure}

Let us dope two fermions at site $x$ in one of the various states
and move one of the fermions from $x$ to site $y$.
The background 8v states change from the original configuration
by the fermion hopping.
For the various state in $|GS;cd\rangle$, one can verify that
movement of doped fermions is restricted to straight lines, i.e., 
the horizontal and vertical lines.
Otherwise $a$ and/or $b$ vertices appear as a result of
fermion hopping (see Fig.\ref{4states}).
In this sense, fermions are deconfined only in one direction.
In the present limit, the fermion system essentially reduces to the 
one-dimensional one, and therefore no SSBs take place as dictated by 
the Coleman theorem. 
This limit is in sharp contrast with Kitaev's point $a=b=c=d=1$.
At Kitaev's point, there are no potential terms of the background
8v configuration and we can simply put $\sigma^3_\ell(x)=1$ at all links.
Then the doped fermions move without any interactions.

\section{Conclusion}
In this paper we have studied behavior of the doped fermions
in the dimer and q8v models.
In most of discussion we employed the IGT description of the models.
We showed that the translational symmetry breaking in the background
dimer or 8v configuration induces SSBs of the internal symmetry
in the doped-fermion sector.
Similar phenomenon is well-known in the quantum chromodynamics 
in which confinement of quarks induces the SSB of the chiral symmetry
of quarks.
Result of the numerical studies on phase structure of the models,
which shows that change of the background configuration really
induces SSBs of the doped fermions, will be reported in the near 
future\cite{IY}.
It is also very interesting to construct more realistic models similar
to the models in this paper for, e.g., strongly-correlated heavy fermions 
and compare phase structure of the models, etc. with experiments.

\newpage

\end{document}